\documentstyle[12pt,epsf]{article}

\begin{document}
\title{ The Implications of the Microwave Background Anisotropies for 
Laser-Interferometer-Tested Gravitational Waves}
\author{L. P. Grishchuk \thanks{Based on a talk presented at the First International LISA Symposium, 9-12 July 1996, Rutherford Appleton 
Laboratory, Oxfordshire, UK} \\
\mbox{\small {Department of Physics and Astronomy}} \\ 
\mbox{\small{ University of Wales, Cardiff CF2 3YB, UK } } \\
\mbox{\small {and}} \\
\mbox{\small{ Moscow State University, 119899 Moscow V234, Russia}}}              
\date{}
\maketitle
\begin{abstract}

The  observed microwave background anisotropies in combination
with the theory of quantum mechanically generated cosmological
perturbations  predict  a  well  measurable  amount  of  relic
gravitational waves in the frequency intervals tested by  LISA
and ground based laser interferometers.

\end{abstract}

At the first glance, the ground-based and space \cite{ben} laser
interferometers for gravity wave observations, as well as the
Weber bar  technique, do not have much in common with the
ongoing and planned radio-astronomical measurements of the
microwave background anisotropies \cite{smo}. The solid-state
detectors are sensitive to gravitational waves in the  $10^3$ Hz
frequency  range,  the  laser interferometers  are  sensitive,
correspondingly, to the frequencies ($10 - 10^3$) Hz and ($10^{-4} -
10^{-1}$) Hz,  whereas  the  microwave  background  anisotropies
directly reflect only the variations in the cosmic temperature
and,  if  they  are caused by gravitational  waves,  can  only
provide  us with information about extremely low-frequency
gravitational  waves  -  ($10^{-16} - 10^{-18}$) Hz and lower \cite{tho}.
However, the basic physics which enables us to see
gravitational waves with the help of laser interferometers  or
microwave background anistropies is exactly the same:
alterations in frequency and phase of an electromagnetic
signal propagating in the field of gravitational waves. Most
importantly, the relic stochastic background of gravitational
waves, to be discussed below, extends from very high
frequencies of the order of $10^8$ Hz to extremely low frequencies
of the order of $10^{-18}$ Hz and lower. It can be observed by all
these techniques, and the predictions about the expected
gravitational  wave amplitudes in various frequency  intervals
are connected to each other by the theory \cite{gri}.

Although the measured large-angular-scale anisotropies in the
cosmic  microwave  background radiation (CMBR) \cite{smo} could be
expected  on  general grounds, their actual  existence raises
certain  theoretical problems.  The observed Universe is far
from  being  homogeneous and isotropic, but becomes more and
more so when one expands the study to larger and larger 
scales. The anisotropies signify the presence in the Universe
of cosmological perturbations with very small amplitudes (the
dimensionless  deviations are of the order of $10^{-5}$) but with
extremely  long wavelengths, of the order of and longer than
the present-day Hubble radius $l_H$, $l_H \approx 2 \times 10^{28}$ cm. 
Regardless of nature of the
perturbations responsible for the observed large-angular-scale
anisotropy,  that is, regardless of whether they are mostly
density perturbations, or rotational perturbations, or
gravitational  waves, or all mixed together, there exists  a
puzzling  question  of  their origin.   The  first  wonder  is
whether  they are remnants of the originally inhomogeneous  and
anisotropic Universe or, alternatively, were generated by some
mechanism in the originally homogeneous and isotropic Universe.
Since  the perturbations of our interest are weak, we can  use
the linearized Einstein equations for the description of their
evolution.

It  is  difficult  to  maintain that these  perturbations  are
simply  survived. The photons of the CMBR have  become  free  and
started their journey to us sometime at the beginning  of  the
matter-dominated  era.  In the preceeding  radiation-dominated
era,  the  general  solution for a Fourier  component  of  the
metric perturbations is \cite{lan}                                   
\begin{equation}
h_{{\bf n}}(\eta, {\bf x}) = A \sin (n\eta + \chi)\frac{1}{a(\eta)}
e^{i {\bf n .x}}
\end{equation}
where  $A $ and $\chi$ are arbitrary constants, $a(\eta)$ is the scale
factor of a FLRW universe
\begin{equation}
  ds^2    =   a^2(\eta)(d\eta^2 - dx^2 - dy^2 - dz^2)
\end{equation}
and  $a(\eta) \sim \eta$ in the radiation-dominated era.  Let us take the
amplitude  $ A $ at the level $A \approx 10^{-5}$, in rough agreement with
observations, and return back in depth of the radiation-
dominated era by sending $\eta$ to zero. The scale factor  $a(\eta)$
diminishes by at least the factor $10^8$ by the time of reaching
the  era of the primordial nucleosynthesis. The dangerous term
$ A \cos n\eta \sin\chi e^{i {\bf n . x}}/a(\eta)$ 
must  be  cancelled out with a great precision, if we do not
want the $h_{{\bf n}}(\eta , {\bf x})$ to become of the order of 1 
and destroy in this  way
the  homogeneity and isotropy of that era.  That is, the phase
$\chi$ must be $0 $ or $\pi$ with a precision of $10^{-3}$, or 
even much higher
if we want to proceed further back in time.  The neighboring
solutions  must all have the same (or differing by $\pi$) phases,
that  is,  the distribution of the phases must be very  narrow
(highly  ``squeezed"); and the waves must be standing, and  not
traveling. (We will see below that all that is automatically
guaranteed if the perturbations are generated quantum
mechanically.) The solution (1) is strictly valid for
gravitational waves, but the same argument is applicable for
density perurbations as well.

It  is  still possible that the perturbations of our  interest
are   classical,   deterministic  remnants   of   a   strongly
inhomogeneous anisotropic universe of a very distant past.  It
is  also possible that, for some miraculous reason, the phases
have been chosen rightly with enormous precision, so that  the
perturbations  are  classical,  deterministic  remnants  of  a
universe  which was almost homogeneous and isotropic from  the
very beginning.  To the present author, these possibilities do
not  seem  to  be  likely, even if they can  be  shown  to  be
consistent  with  all available data.   We  need  to  turn  to
posibilities of generating the perturbations in an  originally
FLRW universe.

In  principle,  there are several options  to  do  that.   For
instance,  one can try to exploit the fact that the number  of
unknown  functions  of  time participating  in  the  perturbed
Einstein  equations  is  always greater  than  the  number  of
equations.   These  functions describe nonadiabatic  pressure,
entropy   perturbations,  anisotropic  stresses,   etc..    By
manipulating  with  these  functions  and  making   additional
assumptions,   such   as   that  these   functions   represent
``cosmological defects" or ``causal seeds", one can produce  the
required  perturbations, but essentially ``by hand".  It  seems
to  the writer that we should first try to build the theory on
a minimal number of hypotheses.

It  is known that the Einstein equations plus basic principles
of  quantum field theory allow (in fact, demand) the  quantum-
mechanical generation of cosmological perturbations  from  the
vacuum  state,  as  a  result  of parametric  (superadiabatic)
interaction   of  the  quantized  perturbations  with   strong
variable gravitational field of the very early Universe.  This
process  is possible for gravitational waves \cite{gri1}, 
for density
perturbations \cite{lu,gri2}, and for rotational perturbations 
\cite{gri3}.
However,  in  the  last two cases we need to assume  that  the
primeval matter was capable of supportng the oscillations  and
that  they  were  properly coupled (similar  to  gravitational
waves)  to  the ``pumping" gravitational  field.   If,  as  is
assumed  in  the  inflationary  hypothesis,  the  very   early
Universe was governed by a scalar field, and if the field  was
minimally   coupled   to  gravity,  the   quantum   mechanical
generation  of  density  perturbations,  in  addition  to  the
inevitable  generation of gravitational waves,  was  possible.
Below,  we  will  follow  the line of  the  quantum-mechanical
generation of cosmological perturbations.

In the presence of perturbations, the metric tensor and  the
energy-momentum tensor can be written in the form

\begin{equation}
ds^2 = a^2(\eta) [d\eta^2 -(\delta_{ij} +h_{ij})dx^idx^j] ,  
\end{equation}

\[
T^{\mu}_{\nu} = T^{\mu (0)}_{\nu} + T^{\mu (1)}_{\nu} .
\]

In  cosmology, one usually considers the $T^{\mu}_{\nu}$ for simple models
of matter, such as perfect fluids or scalar fields. The
perturbations $h_{ij}$ and $T^{\mu (1)}_{\nu}$ 
are linked together by  a  set  of
the linearized Einstein equations.  It is convenient to expand
the  perturbations over spatial harmonics $e^{i{\bf n . x}}$, 
$e^{-i{\bf n . x}}$, where \newline 
${\bf n } = (n^1, n^2, n^3)$ is arbitrary wavevector and the wavenumber $n$
is \newline 
$n  =  [(n^1)^2 + (n^2)^2 +(n^3)^2]^{1/2}$. 

Demonstrating certain  mathematical  skill  and
understanding  of  the physical side of the problem,  one  can
show  that,  for  the simple models of matter,  the  perturbed
Einstein  equations for each $ {\bf n}$-mode and for each of the  three
types  of  cosmological perturbations (density  perturbations,
rotational perturbations, gravitational waves) can be  reduced
to a single differential equation of second order.  It is only
this  one equation that defines the dynamical content  of  the
problem and needs to be solved.  All the components $h_{ij}$, 
$T^{\mu (1)}_{\nu}$ 
can  then  be  found  from  its  solutions  by  algebraic  and
differentiation/integration operations.  This equation has the
form  of  the  equation  for  an oscillator  with  a  variable
frequency.   The frequency varies due to the presence  of  the
time-dependent scale factor $a(\eta)$ which plays the role  of  the
gravitational  pump  field.  The very form  of  this  equation
explains   why   the   cosmological   perturbations   can   be
parametrically  amplified (if the initial classical  amplitude
was not zero) or quantum-mechanically generated from the zero-
point  quantum  oscillations. In case of gravitational  waves,
this equation is

\begin{equation}
\mu^{''}_n + [n^2 -\frac{a^{''}}{a}]\mu_n =0
\end{equation}                                              
and $h_{{\bf n}}(\eta, {\bf x}) \sim \frac{1}{a}\mu_n e^{i {\bf n . x}}$.
 
We will briefly summarize the main points of the quantum-
mechanical generation of cosmological perturbations (for  more
details,  see a recent paper \cite{gri4} and references therein).  The
quantum-mechanical operator for $h_{ij}(\eta,{\bf x})$ can be written as

\begin{equation}
h_{ij}(\eta, {\bf x}) = \frac{C}{a(\eta)}\frac{1}{(2\pi)^{3/2}}
\int_{-\infty}^{\infty} d^3 {\bf n}\sum_{s=1}^{2}
{\stackrel {s} {p}}_{ij}({\bf n}) \frac{1}{\sqrt{2n}} 
[{\stackrel {s} {c}}_{{\bf n}}(\eta)e^{i {\bf n . x}} + 
{\stackrel {s} {c}}_{{\bf n}}^{\dagger} (\eta)e^{-i {\bf n . x}}] . 
\end{equation}     

Each  of the three types of perturbations has two polarisation
states $(s = 1,2)$ described by two polarisation tensors  
${\stackrel {s} {p}}_{ij}$. The
normalization constant $C$ is, up to a numerical factor slightly
different  for  each  type  of the perturbations,  the  Planck
length  $l_{Pl}$,  $l_{Pl}=(G\hbar/c^3)^{1/2}$. 
For gravitational  waves, $C=\sqrt{16\pi} l_{Pl}$.
The time dependent creation and annihilation operators 
$ {\stackrel {s} {c}}_{{\bf n}}^{\dagger} (\eta)$, 
${\stackrel {s} {c}}_{{\bf n}} (\eta)$ 
are governed by the Heisenberg equations of motion. The
parametric nature of the interaction Hamiltonian allows one to
apply  the Bogoliubov transformation and to express the  
${\stackrel {s} {c}}_{{\bf n}}^{\dagger}(\eta)$, ${\stackrel {s} {c}}_{{\bf n}}(\eta)$
 in terms of their initial values 
${\stackrel {s} {c}}_{{\bf n}}^{\dagger}(0)$, ${\stackrel {s} {c}}_{{\bf n}} (0)$. 
The operators ${\stackrel {s} {c}}_{{\bf n}}^{\dagger} (0)$, 
${\stackrel {s} {c}}_{{\bf n}} (0)$ define the
initial  vacuum  state $ | 0 \rangle $ for each ${\bf n}$ and $s$: 
$ {\stackrel {s} {c}}_{{\bf n}} | 0 \rangle = 0$. The field  operator
(5) acquires the form

 \begin{equation}
h_{ij}(\eta, {\bf x}) = \frac{C}{(2\pi)^{3/2}}\int_{-\infty}^{\infty} 
d^3 {\bf n} \sum_{s=1}^{2} {\stackrel {s} {p}}_{ij}({\bf n}) 
\frac{1}{\sqrt{2n}}[{\stackrel {s} {h}}_{n}(\eta) e^{i {\bf n . x}}
{\stackrel {s} {c}}_{{\bf n}}(0) + {\stackrel {s} {h}}_{n}^{\star}(\eta)
e^{-i {\bf n . x}}{\stackrel {s} {c}}_{{\bf n}}^{\dagger} (0)] 
\end{equation}
where ${\stackrel {s} {h}}_{n}(\eta)$ are essentially the solutions to the  single  second-
order  differential equation (for each type of  perturbations)
mentioned above.

In  the  Schrodinger picture, the initial vacuum state 
$| 0_{{\bf n}} \rangle | 0_{- {\bf n}} \rangle $, for
every pair $ {\bf n}, {\bf -n}$ of modes, evolves into a two-mode  squeezed
vacuum quantum state.  The modes affected by the amplification
process will have a large mean value of the number operator $N$
(large  occupation  number) and a large variance of $N$. The
conjugate variance of phase will be highly squeezed near the
values $\chi = 0, \pi$ (in the representation (1)). Classically, the
generated perturbations can be treated as a stochastic
collection of standing waves.

The scale factor $a(\eta)$ of the very early Universe (well before
the era of primordial nucleosynthesis) is  not  known.   It
depends on the  unknown equation of state of  the  extremely
dense matter (we are not quite sure even about the equation of
state in cores of neutron stars). However, it is likely  that
the evolution was significantly different from  the  law  of
expansion of a radiation-dominated universe. If so,
gravitational waves must have been generated, and density
perturbations, as well as rotational perturbations, could  be
generated too.

Certain   properties  of  the  generated   perturbations   are
universal, independent of a concrete form of $a(\eta)$ in the  very
early  Universe,  which we still keep quite arbitrary.   These
properties  are determined by the fact that the  perturbations
are  placed  in  the  squeezed  vacuum  quantum  states.   For
instance,  the  expected (mean) quantum  mechanical  value  of
$h_{ij}(\eta, {\bf x})$ is zero in every spatial point and 
at every moment of time: $\langle 0 |h_{ij}(\eta , {\bf x})|0 \rangle =0$. 
However, the variance is not zero and does depend  on
time:

\begin{equation}
\langle 0 | h_{ij} (\eta , {\bf x})h^{ij}(\eta, {\bf x})| 0 \rangle\hspace{0.1cm} =
\hspace{0.1cm} \frac{C^2}{2\pi^2}\int^{\infty}_0 n \sum_{s=1}^{2} | 
{\stackrel {s} {h}}_n (\eta) |^2 dn .
\end{equation}
Eq. (7) defines the power spectrum

\begin{equation}
P(n) = \frac{C^2}{2\pi^2}n \sum_{s=1}^{2} 
|{\stackrel {s} {h}}_{n} (\eta) |^2 .
\end{equation}                         
For numerical estimates, it is convenient to use the
characteristic amplitude $h(n)$ defining this amplitude as the
standard  deviation (square root of variance) per logarithmic
frequency interval:

\begin{equation}                                         
h(n)= \left [ \frac{C^2}{2\pi^2} n^2 \sum_{s=1}^{2} 
| {\stackrel {s} {h}}_n (\eta)|^2 \right ]^{1/2} .
\end{equation}

It follows from Eq. (8) that for perturbations with
wavelengths shorter than the Hubble radius, the power spectrum
$P(n)$ is not a smooth but an oscillating function of the
frequency (wave number) $n$. Specifically for short
gravitational waves, the generated stochastic background of
the waves is not a stationary but a nonstationary noise, in
the sense that the temporal correlation function of the field
should depend on individual moments of time, and not only on
the time difference. The large variance of $N$ and hence  the
large variance of the amplitude of the perturbations will lead
to large variations in the angular correlation function for
the microwave background anisotropies, and so on. However, in
this presentation, we will not go into the details of
statistics and modulated spectra. We will be happy to show
that we can get a right numerical level of the expected
signal.

The amplitudes and spectra of the generated perturbations
depend on the strength and variability of the pump field,
which in our case is completely determined by the scale factor
$a(\eta)$. Having the freedom of playing with the unknown part  of
$a(\eta)$ describing the very early Universe, one can derive, with
the  degree of completeness allowed by quantum theory, all the
characteristics of the perturbations in the present  Universe.
For instance, some of the gravity-wave spectra derived in this
manner are shown in Fig. 1 (adopted from \cite{gri} and updated).
The original graph used, in the radio-astronomical fashion,
the spectral flux density $F_{\nu}$ (ergs / sec cm$^2$ Hz ster) as its
vertical axis.  In Fig. 1, we have used the vertical axis in
terms of the dimensionless characteristic amplitude $h(\nu)$,  Eq.
(9), where the dimensionless frequency $n$ has been translated
into the present day frequency $\nu$ measured in Hz. The spectral
cosmological $\Omega$-parameter due to the contribution of
gravitational waves, $\Omega_g (\nu)$, is defined as \cite{gri}

\begin{equation}
 \Omega_g (\nu)=\frac{\epsilon_g(\nu)}{\epsilon_{cr}}=
h^2(\nu) \left ( \frac{\nu}{\nu_H} \right )^2
\end{equation}                                   
where $\epsilon_{cr}$ is the critical cosmological energy density and 
$\nu_H \approx 10^{-18}$ Hz is the Hubble frequency. 
Because the graph is the
same, the vertical axis is now not universally homogeneous.

In Fig. 1 one can also see the observational upper limits
for stochastic gravitational waves, marked by arrows, that
were valid at that time, 10 years ago, as well as the expected
sensitivities of then proposed new gravity-wave detectors.
Some of the sensitivity curves, in particular for the space
interferometer now called LISA, should be significantly
modified, but we leave them in the old shape because the
modifications will not be very important for our further
discussion.

\begin{figure}
\centering \epsffile{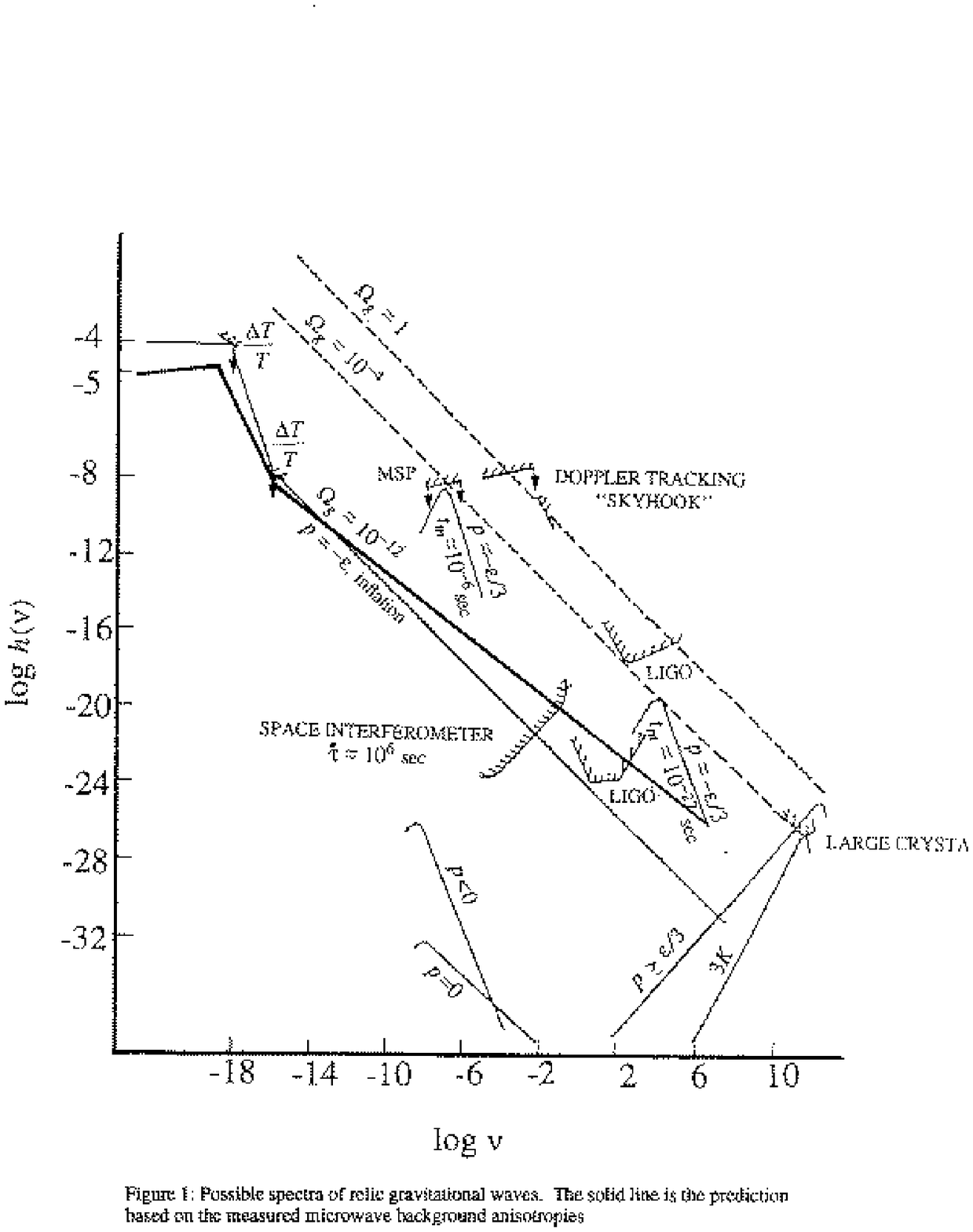}
\end{figure}

Let us start from the two spectra with the maxima near $10^{-8}$ Hz
and $10^{3}$ Hz. These spectra were derived from bizarre
cosmological models \cite{gri} designed specially to produce the
maximum possible amount of gravitational waves in the
frequency intervals where the millisecond pulsar (MSP) and bar-
detector techniques, then most favorite, were operating. One
can also make equally bizarre assumptions about localized
sources (such as colliding bubbles, phase transitions, string
loops, strings with attached monopoles at the ends, etc.) and
produce maxima virtually in any frequency interval, for the
benefit of every individual experimental group. Generally
speaking, we should keep eyes open to all these possibilities,
they  all are not forbidden.  A different question is  whether
we  will  be surprised if the predicted signal is not detected
and what we will learn from that fact.

As was explained above, the quantum-mechanical (parametric)
generating mechanism relies only on the validity of general
relativity and basic principles of quantum field theory. The
law of expansion of the very early Universe is not known but
this is what we will learn, or at least will place
restrictions on, by detecting or not detecting the predicted
signal. For instance, the once popular cosmological model
governed  by matter with the stiff equation of state $p=\epsilon$ can
already  be ruled out, because the amount of the created 
high-frequency gravitons
would  be  too  big and would be inconsistent  with  available
cosmological data \cite{gri1,zel}.

A large class of expanding cosmological models is described by
the scale factor 
\begin{equation}
a(\eta)=l_0 | \eta |^{1+\beta}
\end{equation}                                             
where $l_0$ and $\beta$ are arbitrary constants. The $\eta$ time grows from
$-\infty$ and $\beta  <  -1$ at the initial stage of expansion. 
The constant $l_0 $ has the dimensionality of
length and is effectively responsible for the Hubble-radius
(time-dependent, unless $\beta = -2$) of the very early Universe.
With this scale factor, the Einstein equations require the
equation of state to be in the form 
\begin{equation}
p=\frac{1-\beta}{3(1+\beta)}\epsilon .
\end{equation}                                              
For  $\beta  =  -2$ one has $p = -\epsilon$, see Eq. (12), the case called
inflation.

Solving the gravity-wave equation (4) one can show that
today's values of the characteristic amplitude $h(\nu)$ should  be
as follows (ignoring the modulation of the spectrum which
takes place for $\nu \gg \nu_H$):
     
For $\nu \leq \nu_H$, 
\begin{equation}
 h(\nu) \approx \frac{l_{Pl}}{l_0} \left ( \frac{\nu}{\nu_H} 
\right )^{\beta +2} . 
\end{equation}                                             

For  $\nu_H  \leq \nu  \leq \nu_m$,  where $\nu_m$ is determined  
by the time of
transition from the radiation-dominated era to the matter-
dominated era, $\nu_m \approx 10^{-16}$ Hz, 
\begin{equation}
 h(\nu) \approx \frac{l_{Pl}}{l_0} \left (\frac{\nu}{\nu_H} 
\right )^{\beta} .
\end{equation}                                             

For  $\nu_m  \leq \nu < \nu_c$, where $\nu_c $ labels the highest frequency
waves marginally affected by the amplification process and
above which the spectrum sharply falls down, $\nu_c \approx 10^8$ Hz in
currently discussed models,

\begin{equation}
h(\nu) \approx \frac{l_{Pl}}{l_0} \left (\frac{\nu_m}{\nu_H} 
\right )^{\beta}
\left (\frac{\nu}{\nu_m} \right )^{\beta +1}
\end{equation}                                        
The scale factors (11) which are power-law dependent on $\eta$ time
generate spectra which are power-law dependent on frequency $\nu$ 
\cite{gri1}.

As  we  see,  the numerical level of the predicted  amplitudes
depends on the fundamental constants $G, c, \hbar$ combined  in $l_{Pl}$
and  a  couple of unknown cosmological parameters, such as  $l_0$ 
and $\beta$.  Of course, it is not for the first time that something
observable is beautifully expressed in terms of the
fundamental constants and a couple of parameters only. For
instance, the maximal masses and radii of white dwarfs and
neutron stars depend essentially only on $G, c, \hbar$, and the
masses of an electron $m_e$ and a baryon $m_B$ (see, for example,
\cite{sha}). In principle, if did not not know $m_e$ and $m_B$, we could
derive them from accurate astronomical measurements!  It seems
to  the  author  that in the case of relic  gravitons  we  are
dealing with the problem of a similar simplicity and deepness.

Let  us continue our review of Fig. 1.  A spectrum, part  of
which  is shown as a horizontal line at the level $h(\nu) = 10^{-4}$,
was  derived  from the $\beta = -2$ model and placed at the  highest
level  allowed  by  then  existing  limits  on  the  microwave
background anisotropy $\Delta T/T$.  It is known \cite{sac} that the long-
wavelength  cosmological  perturbations produce the large-
angular-scale anisotropy $\Delta T/T $ of the order of 
$\Delta T/T \approx h$. The
upper limit of that time $\Delta T/T < 10^{-4}$ required
  $h  <  10^{-4}$. The horizontal position of this  part  of  the
spectrum (use Eq. (13) for $\beta = -2$) explains why this spectrum
is  called ``flat" (or Harrison-Zeldovich) spectrum.  All waves
with  present frequencies $\nu > \nu_H$ were ordered long ago in  the
``flat"  spectrum, but adiabatically decreased their amplitudes
by now.  In particular, if $\beta = -2$ and $\nu > \nu_m$ one has 
$h(\nu) \sim  \nu^{-1}$ \cite{sta}. The full present-day spectrum is a 
continuation of the
horizontal line to higher frequencies, as shown in Fig. 1.
As was already emphasized above, we ignore oscillations in the
power  spectrum $P(n)$ of the $h$-field itself.  As for the  power
spectrum  of  the  energy density $\epsilon_g$, it is expected to be
smooth, because the $\sin^2 (n\eta + \chi)$ 
oscillations in $h^2$ combine with the $\cos^2(n\eta +\chi)$ 
oscillations  in  $(h')^2$ to produce a smooth function for the
sum. The $\beta = -2$ model and the $\Delta T/T$ limits of that time did
not allow $\Omega_g$ in the frequency bands $\Delta\nu \approx \nu$ 
to be larger than
$10^{-12}$ for all ground-based and space techniques. (The old
graph for the space interferometer sensitivity does still
appear promising for detecting such a signal but it is now
known to be overly optimistic \cite{ben}.)

The situation has considerably changed in the recent years.

First,  the  large-angular-scale anisotropy has been  actually
detected \cite{smo}, so we are now dealing with the detected  signal
$\Delta T/T \approx 10^{-5}$, and not with the upper limit 
$\Delta T/T < 10^{-4}$.  A very
important question is whether we can attribute, say, a half of
the detected signal to gravitational waves, assuming of course
that  the  $\Delta T/T $ is  caused by cosmological perturbations  of
quantum-mechanical origin, as we argued in the very beginning
of this paper. Without being able presently to answer this
question observationally, we should rely on the theory. The
theory definitely says  yes. A possible contribution of
quantum mechanically generated density perturbations can be
of the same order of magnitude as (in fact, somewhat smaller
than) the gravity-wave contribution, but cannot be much
higher \cite{gri2}.  Specifically for models with the scale factors
(11)  governed by a scalar field, the characteristic amplitude
of  the  long-wavelengths metric perturbations $h(\nu)$ associated
with  the  density perturbations and responsible for  
$\Delta T/T $ is
described  by  exactly the same formula as  formula  (13)  for
gravitational waves.  This is not surprising since  the  basic
dynamical  equation for the scalar field density perturbations

\[
\mu''_n + [n^2 -\frac{(a\sqrt{\gamma})''}{a\sqrt{\gamma}}]\mu_n = 0 
\]
(where $\gamma \equiv 1+ (a/a')'$ and the scalar field potential is 
arbitrary) 
is  not  only similar, but is exactly the same as Eq. (4)  for
gravitational waves, when the scale factor $a(\eta)$ at the initial
stage of expansion is taken as a power-law function (11) (in which 
case, $\gamma =$ const. and $\gamma$ drops out of the equation). The
numerical coefficients in formula (13) are somewhat  in  favor
of  gravitational waves, but it is not numerical  coefficients
of  order 1 that we are now discussing.  What is important for
us  is  that  the  observed $\Delta T/T$ can now be taken  as  an
experimental point for relic gravitational waves.  If we  have
still used the $\beta = -2$ model, the entire inflationary graph  of
Fig.  1  should have been shifted down by 1 order of magnitude
in terms of $h(\nu)$, predicting a hopelessly small 
$\Omega_g \approx 10^{-14}$ for
frequencies of our interest, where direct measurements of  the
gravity-wave background are possible.

Second,  the  processing of the COBE data has  allowed  us  to
obtain  some information \cite{be,bru} about the power-law spectral
index of primordial perturbations. Usually, the COBE
data  are processed under the assumption that the anisotropies are caused
by density perturbations. Fortunately, it does not matter what
one  thinks about the perturbations while processing the data:
the  effects  of gravitational waves and density perturbations
on  the  large-angular-scale anisotropies are about  the  same
when amplitudes and spectral indexes ($\beta +2$ in our notation)  of
the  corresponding metric perturbations are the same, see  Eq.
(13).

Density perturbations with the Harrison-Zeldovich spectrum are
often desribed by the spectral index $n = 1$ (this $n$ should  not
be  confused with the dimensionless wave number $n$ used in this
paper).   The relationship between this spectral index $n $ and
the parameter $\beta$ is 
\begin{equation}
n \equiv 2 \beta + 5. 
\end{equation}                                                  
Information about the primordial spectral index extracted from
the  large-angular-scale  anisotropies  is  information  about
gravitational waves (when they dominate or provide, at  least, 
a  half  of  the signal) even if one thinks that the processed
anisotropies are entirely produced by density perturbations.

The authors of \cite{be} concluded:  ``...we  find  a  power-law
spectral  index of $n = 1.2 \pm 0.3$ ...", ``The power  spectrum  of
the  COBE  DMR  data  is consistent with [a  
Peebles-Harrison-Zeldovich  $n  =  1$  universe]". The authors of \cite{bru}, 
who
processed  the  same  set of data but in a  different  manner,
concluded:  ``The spectral parameter of the power  spectrum  of
primordial perturbations $n = 1.84 \pm 0.29$ [is] estimated", ``The
power spectrum estimation results are inconsistent with the
Harrison-Zeldovich $n = 1$ model with the confidence \newline 
99 \%".

It is difficult for us to judge whether the $n = 1 $ model is
ruled out at the confidence level 99 \%, according to \cite{bru}, or
at the confidence level 60 \%  or so, according to \cite{be}.
However, in these two results, we see the indication that the
spectral index $n $ is indeed larger than 1. As a compromise, we will first
take the value $n = 1.4$ and derive its consequences for
gravitational waves.

The  spectral index $n = 1.4$ translates into $\beta = -1.8$, see 
Eq. (16). The  $\nu < \nu_H$ part of the spectrum is not  any  longer
``flat"  but  gives  more power to higher frequencies,  
$h(\nu)  \sim  \nu^{0.2}$,  see  Eq.  (13).  The value 
$h(\nu_H) =  10^{-5}$  is  fixed  by
observations.  The number $\beta = -1.8$ should also be used in  
Eq. (14) and Eq. (15). This gives $h(\nu_m) =  10^{-8.6}$, a 
slightly smaller amplitude than $h(\nu_m) = 10^{-8}$ for 
the $\beta  =  -2$ model.
But for the higher frequencies the results are very
encouraging.  At the LISA-tested frequency $\nu = 10^{-3}$ Hz we get
$h  = 10^{-19}$ and $\Omega_g = 10^{-8}$. At the LIGO/VIRGO-tested frequency 
$\nu=  10^2 $ Hz  we get $h = 10^{-23}$ and $\Omega_g = 10^{-6}$. To plot 
the  full
spectrum, one can essentially shift down the entire graph of the 
$\beta = -2$ model and  slightly rotate the graph around  the  fixed
observational point $h = 10^{-5}$ at $\nu = \nu_H$. This produces a solid
line shown in Fig. 1. This line is no longer a result of a
simple model-building, it is now supported, at least
partially, by observations. The fact that the graph consists
of straight line pieces meeting at corners is accounted for by
the  nature  of  our approximation: strictly  power-law  scale
factors  joined at the transition points between the  initial,
radiation-dominated, and matter-dominated  eras.   At  a  more
accurate graph, the corners will be rounded and lines will  be
slightly bent.

We  have  to  admit  that  it  is  not  so  easy  to  give   a
``microphysical" explanation to the derived parameter $\beta = -1.8$.
This  value of $\beta$ corresponds to the equation of state
$p = -1.2 \,\, \epsilon$, see Eq. (12). With the equation of state of 
this  kind, energy density of the matter, as well as curvature of the 
space-time, increase in course of expansion.  However, the value 
$\beta =-1.8$ is marginally consistent with the assumption that the
Planck densities are not encountered in course of the
evolution \cite{gri5}. The scalar fields, often considered in the
context of the inflationary hypothesis, are only capable  of
providing an equation of state with $\beta \leq -2$. Possibly, a
solution to the ``microphysical" side of the problem can be
found along the lines of the ``superstring motivated"
cosmologies \cite{gas}.

The direct detection of the useful noise (a stochastic gravity-
wave  signal) by noisy detectors can be achieved with the help
of  a  standard technique of cross-correlating the outputs  of
two  or  more detectors \cite{gri6,mi,com}. The LISA is not planned
to have two independent detectors \cite{ben}. However, the predicted
signal is so high, $\Omega_g = 10^{-8}$ at $\nu = 10^{-3}$ Hz and 
$\Omega_g = 5 \times 10^{-8}$ at
$\nu  =  10^{-1}$  Hz, that one can probably recognize the signal  by
comparing   the   observational  data  with   the   calculated
sensitivity  of  the instrument. Fortunately,  at  frequencies
around  $10^{-2}$  Hz   and higher, the contaminating  gravity-wave
noise  from  compact  binaries is expected  to  be  below  the
projected LISA sensitivity. It is important to note that the conservative
value $n = 1.2$ ($\beta = -1.9$) of the spectral index does still lead to a
quite well measurable signal: $h = 10^{-20.5}$, $\Omega_g = 10^{-11}$ 
at $\nu = 10^{-3}$ Hz and $h = 10^{-25}$, $\Omega_g = 10^{-10}$ at
$\nu = 10^2$ Hz.

The  first  evaluation \cite{gri6} of a possibility to  detect  relic
gravitons  by a cross-correlating technique was based  on  the
assumptions  that the flux density behaves as $F_{\nu} \sim \nu^{-1}$  (that
is,
$h(\nu)  \sim \nu^{-1}$, like in the $\beta = -2$ 
model considered above), that
the  expected amplitude is at the level (in current  notation)
of $\Omega_g = 10^{-4}$, and that the electromagnetic detectors operating
in the high-frequency band $\nu = 10^7$ Hz are being used.  At that
time,   these   assumptions  were  of  kind  of  a   stretched
imagination.  Presently, this possibility may turn out  to  be
more  realistic  in  view of the fact that $\Omega_g(\nu)$ is growing
toward  the  higher frequencies, see Eq. (15) and Fig. 1.
However, the prospect for the high-frequency techniques,  such
as  bars and electromagnetic detectors, may be not as good  as
it  looks  on the graph.  The high-frequency parts of  spectra
generated  in  simple models of the very early  Universe  have
tendency to deviate down from the straight line corresponding
to a strict power-law dependence (11). Generally speaking, the
farther  we  extrapolate the spectrum  from  the  experimental
point  at $\nu = \nu_H$, the less confident we are.  In this respect,
the   LISA  has  an  additional  advantage  of  operating   at
relatively low frequencies.

So, what is my conclusion? Urgently fly LISA!

\end{document}